# MAGNETIC MOMENTS OF $(J^P = 3/2^+)$ HEAVY BARYONS USING EFFECTIVE MASS AND SCREENED CHARGE SCHEME

Rohit Dhir and R.C. Verma

*Department of Physics,*

*Punjabi University, Patiala-147002, India*

Email: dhir.rohit@gmail.com, rcverma@gmail.com.

## ABSTRACT

Magnetic moments of heavy charmed baryons with $J^P = 3/2^+$ are predicted employing the concept of effective quark mass and screened charge of quark. We also extend our scheme to predict the $3/2^+ \to 1/2^+$ transition magnetic moments. A comparison of our results with the predictions obtained in recent models is presented.

PACS number(s): 13.40.Em, 14.20.Lq, and14.65.Dw

## 1. INTRODUCTION

The recent predictions of the heavy baryon properties have become a subject of renewed interest due to the experimental facilities at Belle, BABAR, DELPHI, CLEO, CDF etc [1-6]. All the charm = 1 spatial-ground-state baryons carrying $J^P = 3/2^+$ have been observed, and their masses have also been measured. Experimentally, there has been great advancement in the measurement of the baryon magnetic moments. There now exist measurements of magnetic moments of all the octet $J^P = 1/2^+$ baryons, except for the $\Sigma^0$ which has a life time too short for it to travel a significant distance even at the higher energies now available [7]. Two of the magnetic moments of baryon decuplet $J^P = 3/2^+$ have also been measured [7-10]. Theoretically, there exist serious discrepancies between the quark model predictions and experimental results [11]. Various theoretical model based on symmetry breaking, non relativistic quark model (NRQM), configuration mixing, the chiral quark model (ChQM), meson cloud corrections, the QCD based quark model with loop corrections, lattice QCD, effective mass scheme, bag model, the Skyrmion model, chiral quark soliton model,chiral perturbation theory etc. [12-26], have been used to remove the discrepancies, but with partial success. Recently, Sum rules have also been used to calculate the magnetic moments and radiative decays of the decuplet and heavy flavor baryons in [27-30]. Earlier, Bains and Verma [31] have obtained an excellent fit of the baryon magnetic moments with experiment using effective quark mass and screened charge of quarks formalism. Later, Kumar, Dhir and Verma [32] extended this formalism to predict the magnetic moments of $J^P = 1/2^+$ charmed baryons. In this paper, we further extend this scheme to predict the magnetic moments of $J^P = 3/2^+$ charm baryons and transition ($3/2^+ \rightarrow 1/2^+$) moments employing the quark effective mass and screened charge of quark inside respective baryons.

## 2. EFFECTIVE QUARK MASS AND CHARM BARYON MAGNETIC MOMENT:

The baryon mass is taken to be the sum of the quark masses plus spin dependent hyperfine interaction [11, 23],

$$M_B = \sum_i m_i + \sum_{i<j} b_{ij} \; \vec{s}_i \cdot \vec{s}_j \, , \tag{1}$$

where,

$$b_{ij} = \frac{16\pi\alpha_s}{9m_i m_j} < \Psi_0 \left| \delta^3(\vec{r}) \right| \Psi_0 > . \tag{2}$$

$\vec{s}_i$ and $\vec{s}_j$ are the spin operators of the $i^{th}$ quark and $j^{th}$ quark, and $\Psi_0$ is the baryon wave function. Following the formalism described in earlier work [32], for $J^P = 3/2^+$ (*aab*)-type baryons we write

$$m_1^{eff} = m_2^{eff} = m + \alpha\, b_{12} + \beta\, b_{13} , \tag{3}$$

$$m_3^{eff} = m_3 + 2\beta\, b_{13} \tag{4}$$

and

$$m_1 = m_2 = m \text{ and } b_{13} = b_{23} . \tag{5}$$

The $\alpha$ and $\beta$ parameters are calculated through,

$$\begin{aligned} M_B = \sum_i m_i^{eff} &= \sum_i m_i + \sum_{i<j} b_{ij}\ \vec{s}_i \cdot \vec{s}_j \\ &= 2m + m_3 + \frac{b_{12}}{4} + \frac{b_{13}}{2}; \end{aligned} \tag{6}$$

yielding

$$\alpha = \beta = \frac{1}{8}., \tag{7}$$

Thus,

$$m_1^{eff} = m_2^{eff} = m + \frac{b_{12}}{8} + \frac{b_{13}}{8}, \tag{8}$$

and

$$m_3^{eff} = m_3 + \frac{b_{13}}{4} . \tag{9}$$

For (*abc*)-type $J^P = 3/2^+$ baryons,

$$m_1^{eff} = m_1 + \frac{b_{12}}{8} + \frac{b_{13}}{8}, \tag{10}$$

$$m_2^{eff} = m_2 + \frac{b_{23}}{8} + \frac{b_{12}}{8}, \tag{11}$$

$$m_3^{eff} = m_3 + \frac{b_{23}}{8} + \frac{b_{13}}{8}. \tag{12}$$

Similarly, for (*aaa*)-type $J^P = 3/2^+$ baryons,

$$m_1^{eff} = m_2^{eff} = m_3^{eff} = m + \frac{b_{12}}{4}, \tag{13}$$

and

$$b_{12} = b_{23} = b_{13}.$$

Values of quark masses and hyperfine interaction terms $b_{ij}$ are obtained from the known iso-multiplet masses $N$, $\Delta$, $\Lambda$ and $\Lambda_c$ ,

$$\begin{aligned} &m_u = m_d = 362 \text{ MeV}, m_s = 539 \text{ MeV}, \; m_c = 1709 \text{MeV}, \\ &b_{uu} = b_{ud} = b_{dd} = 196 \text{ MeV}, \end{aligned} \tag{14}$$

which in turn yield :

$$\begin{aligned} b_{us} &= b_{ds} = \left( \frac{m_u}{m_s} \right) b_{uu} = 132\, MeV, \\ b_{ss} &= \left( \frac{m_u}{m_s} \right)^2 b_{uu} = 89\, MeV, \\ b_{uc} &= b_{dc} = \left( \frac{m_u}{m_c} \right) b_{uu} = 42\, MeV, \\ b_{sc} &= \left( \frac{m_u^2}{m_s\, m_c} \right) b_{uu} = 28\, MeV, \\ b_{cc} &= \left( \frac{m_u}{m_c} \right)^2 b_{uu} = 9\, MeV. \end{aligned} \tag{15}$$

Using these values of quark masses and hyperfine interaction terms $b_{ij}$ obtained, we calculate effective quark masses for $J^P = 3/2^+$ as follows:

i. For (C = 0) decuplet baryons,

$$\begin{aligned} &m_u^{\Delta} = m_d^{\Delta} = 411 \text{MeV}, \\ &m_u^{\Sigma^*} = m_d^{\Sigma^*} = 403 \text{MeV}, m_s^{\Sigma^*} = 572 \text{MeV}; \\ &m_s^{\Omega} = 561 \text{MeV}; \\ &m_u^{\Xi^*} = m_d^{\Xi^*} = 395 \text{MeV}, \\ &m_s^{\Xi^*} = 567 \text{MeV}. \end{aligned} \tag{16}$$

ii. For (C = 1) sextet baryons,

$$\begin{aligned}
&m_u^{\Sigma_c^*} = m_d^{\Sigma_c^*} = 392\,\text{MeV},\\
&m_c^{\Sigma_c^*} = 1720\,\text{MeV};\\
&m_u^{\Xi_c^*} = m_d^{\Xi_c^*} = 384\,\text{MeV},\\
&m_s^{\Xi_c^*} = 559\,\text{MeV},\ m_c^{\Xi_c^*} = 1718\,\text{MeV};\\
&m_s^{\Omega_c^*} = 554\,\text{MeV},\ m_c^{\Omega_c^*} = 1716\,\text{MeV}.
\end{aligned} \tag{17}$$

iii. For (C = 2) triplet baryons,

$$\begin{aligned}
&m_u^{\Xi_{cc}^*} = m_d^{\Xi_{cc}^*} = 373\,\text{MeV},\\
&m_c^{\Xi_{cc}^*} = 1715\,\text{MeV};\\
&m_s^{\Omega_{cc}^*} = 546\,\text{MeV},\ m_c^{\Omega_{cc}^*} = 1715\,\text{MeV}.
\end{aligned} \tag{18}$$

iv. For (C = 3) singlet baryon,

$$m_c^{\Omega_{ccc}^*} = 1711\,\text{MeV}. \tag{19}$$

## 3. MAGNETIC MOMENTS OF $(J^P = 3/2^+)$ BARYONS

In the present scheme, magnetic moments of $J^P = 3/2^+$ baryons are obtained by sandwiching the following magnetic moment operator [13] between the appropriate baryon wave functions:

$$\vec{\mu} = \sum_i \mu_i^{eff} \vec{\sigma}_i, \tag{20}$$

where

$$\mu_i^{eff} = \frac{e_i}{2m_i^{eff}}, \quad \text{for } i = u, d, s \text{ and } c.$$

Expressions for magnetic moments of $J^P = 3/2^+$ baryons are given in the Tables 1 and 2. Using these effective quark masses, we calculate the magnetic moments of $J^P = 3/2^+$ baryons as given in column 3 of Table 2. For the sake of comparison we also give the results of the Naïve quark model and different models namely QCD sum rules [28], light cone QCD sum rules [29], hyper central model [33,34], non-relativistic quark model [35] and chiral quark-soliton model [36]. Obtained magnetic moments are small as compared to the naïve quark model results because of large effective mass of the quarks inside the baryons.

## 4. TRANSITION ($3/2^+ \to 1/2^+$) MOMENTS

Proceeding in the same manner as in last section, we obtained transitions ($3/2^+ \to 1/2^+$) moments by sandwiching (20) between the appropriate $3/2^+$ and $1/2^+$ baryon wave functions. In order to evaluate $B_{3/2} \to B_{1/2}\,\gamma$ transition magnetic moments, we take geometric mean of effective quark masses of the constituent quarks of initial and final state baryons

$$m_i^{B_{3/2}B_{1/2}} = \sqrt{m_i^{B_{3/2}} m_i^{B_{1/2}}}\ , \tag{21}$$

where $m_i^{B_{1/2}}$ is the effective quark mass of $i^{th}$ quark inside $J^P = 1/2^+$ baryon. These masses are obtained from the following relations derived in the earlier work [32]:

i. For (*aab*)-type baryons,

$$m_1^{eff} = m_2^{eff} = m + \frac{b_{12}}{8} - \frac{b_{13}}{4} \quad \text{and} \quad m_3^{eff} = m_3 - \frac{b_{13}}{2}. \tag{22}$$

ii. For (*abc*) $\Lambda^0$-type baryons,

$$m_1^{eff} = m_2^{eff} = m - \frac{3b_{12}}{8} \quad \text{and} \quad m_3^{eff} = m_3\,. \tag{23}$$

iii. For (*abc*) $\Sigma^0$-type baryons,

$$m_1^{eff} = m_1 + \frac{b_{12}}{8} - \frac{b_{13}}{4}, \tag{24}$$

$$m_2^{eff} = m_2 + \frac{b_{23}}{8} - \frac{b_{12}}{4}, \tag{25}$$

$$m_3^{eff} = m_3 - \frac{b_{23}}{4} - \frac{b_{13}}{4}. \tag{26}$$

Using these relations we obtain the effective quark masses for $J^P = 1/2^+$ baryons as follows:

i. For (C=0) octet baryons,

$$m_u^p = m_d^n = 337\,\text{MeV}, m_u^p = m_d^n = 264\,\text{MeV};$$
$$m_u^\Sigma = m_d^\Sigma = 354\,\text{MeV},\ m_s^\Sigma = 473\,\text{MeV};$$

$$m_u^\Lambda = m_d^\Lambda = 289\,\text{MeV},\ m_s^\Lambda = 539\,\text{MeV};$$

$$m_u^{\Xi} = m_d^{\Xi} = 296\,\text{MeV},\ m_s^{\Xi} = 517\,\text{MeV}; \tag{27}$$

ii. For (C=1) anti-triplet baryons,

$$\begin{aligned} &m_u^{\Lambda_c} = m_d^{\Lambda_c} = 289\,\text{MeV}, m_c^{\Lambda_c} = 1709\,\text{MeV}; \\ &m_u^{\Xi_c} = m_d^{\Xi_c} = 313\,\text{MeV}, \\ &m_s^{\Xi_c} = 490\,\text{MeV},\ m_c^{\Xi_c} = 1709\,\text{MeV}; \end{aligned} \tag{28}$$

iii. For (C=1) sextet baryons,

$$\begin{aligned} &m_u^{\Sigma_c} = m_d^{\Sigma_c} = 376\,\text{MeV}, m_c^{\Sigma_c} = 1688\,\text{MeV}; \\ &m_u^{\Xi_c'} = m_d^{\Xi_c'} = 368\,\text{MeV}, \\ &m_s^{\Xi_c'} = 549\text{MeV},\ m_c^{\Xi_c'} = 1692\,\text{MeV}; \\ &m_s^{\Omega_c} = 543\,\text{MeV},\ m_c^{\Omega_c} = 1695\,\text{MeV}. \end{aligned} \tag{29}$$

iv. For (C=2) triplet baryons,

$$\begin{aligned} &m_u^{\Xi_{cc}} = m_d^{\Xi_{cc}} = 341\,\text{MeV}, \\ &m_c^{\Xi_{cc}} = 1700\,\text{MeV}, \\ &m_s^{\Omega_{cc}} = 525\,\text{MeV},\ m_c^{\Omega_{cc}} = 1703\,\text{MeV}. \end{aligned} \tag{30}$$

Obtained numerical values of the transition magnetic moments of charmless and charmed baryons are given in column 3 of Tables 4. In order to compare, we also present the results of the Naïve quark model and other models namely QCD sum rules [28] and light cone QCD sum rules [30]. Because of large effective masses of the quarks inside the baryons, our results are small as compared to the naïve quark model values.

## 5. MAGNETIC MOMENTS WITH EFFECTIVE MASS AND SHIELDED QUARK CHARGE:

In addition to the variation of the quark mass, the modification in the magnetic moment of a quark due to its environment may also occur through its effective charge. For example, the charge of the quark, when probed by a photon, may be screened due to the presence of the neighboring quarks [37]. It is in some sense similar to the shielding of the nuclear charge of the helium atom due to its outer electron cloud. We take the effective charge to depend linearly on the charge of the shielding quarks. Thus effective charge of quark, $a$, in the baryon $B$ ($a$, $b$, $c$) is taken as [37]

$$e_a^{B} = e_a + \alpha_{ab} e_b + \alpha_{ac} e_c \ , \tag{31}$$

where $e_a$ is the bare charge of quark $a$. Taking $\alpha_{ab} = \alpha_{ba}$ and invoking the isospin symmetry, we obtain the following constraints :

$$\begin{aligned} &\alpha_{uu} = \alpha_{ud} = \alpha_{dd} = \beta, \\ &\alpha_{us} = \alpha_{ds} = \alpha, \\ &\alpha_{ss} = \gamma, \\ &\alpha_{uc} = \alpha_{dc} = \beta', \\ &\alpha_{sc} = \delta, \\ &\alpha_{cc} = \gamma' . \end{aligned} \tag{32}$$

Using the SU(3) flavor symmetry, we reduce them to three, as given below.

$$\begin{aligned} &\alpha = \beta = \gamma \ , \\ &\beta' = \delta \ , \end{aligned} \tag{33}$$

and SU(4) flavor symmetry further reduces them to a single parameter through

$$\gamma = \gamma' = \delta. \tag{34}$$

Redefining the magnetic moment operator,

$$\vec{\mu} = \sum_i \frac{e_i^B}{2m_i^{eff}} \vec{\sigma}_i \tag{35}$$

we determine the baryon magnetic moments. Using the $p$, $n$ and $\Lambda$ moments as input, we fix the quark masses for numerical calculations, $m_u = m_d = 370$ MeV, $m_s = 494$ MeV, and $\alpha = 0.033$. Here we keep $m_c = 1680$ MeV. The obtained numerical values are given in column 4 of Tables 3 and 4.

## 6. NUMERICAL RESULTS AND DISCUSSION

Presently, only two experimental measurements of magnetic moments are available for comparison. Our results for $\mu_{\Delta^{++}}$ are consistent with experimental value $4.52 \pm 0.95$

n.m. [8] and are well within the experimental range given by the particle data group [7]. Note that $\mu_{\Delta^{++}}$ has decreased from the naïve quark model 5.58 to 4.56 n.m. in effective mass scheme, coming closer to the experiment. In effective mass scheme $\mu_{\Omega^-}$ =1.67 n.m. less than experimental value however when effect of screened charge is included it improves in the right direction i.e. $\mu_{\Omega^-}$ =1.90 n.m., which is consistent with one of the experimental value [9] within errors. In case of charm baryons, there is no experimental information available about the magnetic moments to compare with. However, the present results are compared with the predictions of other models [29, 33, 35]. Our predictions of magnetic moments of $J^P = 3/2^+$ charmed baryons are comparable to the predictions based on hyper central model [33] and NRQM [35]. T. M. Aliev *et al.* [29] calculated, $\mu_{\Sigma_c^{*++}}$ = (4.81 ± 1.22) n.m., $\mu_{\Sigma_c^{*+}}$ = (2.00 ± 0.46) n.m., $\mu_{\Sigma_c^{*0}}$ = (-0.8 ± 0.2) n.m., which are consistent with our results. We wish to point out that all of our results are small as compared to naïve quark model, roughly the same is also observed for the other models.

On the other hand, out of all the $3/2^+ \to 1/2^+$ baryon transition magnetic moments only one experimental value is known i.e. $\mu_{\Delta^0 \to n}$ = 3.23 ± 0.10 n.m. [8, 26]. However, we obtain $\mu_{\Delta^0 \to n}$ = 2.58 n.m. in effective mass scheme which is quite small as compared to the experimental value. It may be noted that inconsistency of $\mu_{\Delta^0 \to n}$ like transition moments with experiment is a long standing problem and none of the model has been able to explain it. The light cone QCD sum rule based predictions [30] of the transition $(3/2^+ \to 1/2^+)$ magnetic moments $\mu_{\Sigma_c^{*++} \to \Sigma_c^{++}}$ = 1.06 ± 0.38 n.m., $\mu_{\Sigma_c^{+} \to \Lambda_c^{+}}$ = 1.48 ± 0.55 n.m., $\mu_{\Xi_c^{*+} \to \Xi_c^{+}}$ = 1.47±0.66 n.m., and $\mu_{\Xi_c^{*0} \to \Xi_c^{0}}$ = 0.16±0.07 n.m. are found to be consistent with our results. However, transition magnetic moments $\mu_{\Sigma_c^{*+} \to \Sigma_c^{+}}$ = 0.45±0.11 n.m. is large and $\mu_{\Sigma_c^{*0} \to \Sigma_c^{0}}$ = 0.19±0.08 n.m. is small as compared to our results. Note that T.M. Aliev, *et al.* [30] have given their results in natural magneton $(e\hbar/2cM_B)$, however to convert to nuclear magneton we multiply the entire magnetic moments with $2\,m_N/(M_{B_{3/2^+}} + M_{B_{1/2^+}})$. In addition to the single charm quark baryon transition magnetic moments we also predict doubly charmed baryon transition magnetic moments.

We wish to remark here that the effective mass scheme has worked well for explaining magnetic moments of $J^P = 1/2^+$ octet baryons [31, 32]. With growing experimental facilities, a few experimental groups (BTeV and SELEX Collaborations) are expected to do measurements in near future, which would test the present scheme for the charm baryons.

## REEFERENCES

## TABLE CAPTIONS

Table 1. Expressions for magnetic moments ($J^P = 3/2^+$) baryons using effective quark masses (in nuclear magneton)

Table 2(a). $\left|\mu_{3/2^+ \to 1/2^+}\right|$ Transition magnetic moments of charmless baryons using effective quark masses (in nuclear magneton)

Table 2(b). Expressions for $\left|\mu_{3/2^+ \to 1/2^+}\right|$ transition magnetic moments of charmed baryons using effective quark masses (in nuclear magneton)

Table 3(a). Magnetic Moments of ($J^P = 3/2^+$) decuplet baryons using effective quark masses (in nuclear magneton)

Table 3(b). Magnetic Moments of charmed ($J^P = 3/2^+$) baryons using effective quark masses (in nuclear magneton)

Table 4(a). $\left|\mu_{3/2^+ \to 1/2^+}\right|$ transition magnetic moments of charmless baryons using effective masses (in nuclear magneton)

Table 4(b). $\left|\mu_{3/2^+ \to 1/2^+}\right|$ transition magnetic moments of charmed baryons using effective masses (in nuclear magneton)

Table 1. Expressions for magnetic moments ($J^P = 3/2^+$) baryons using effective quark masses (in nuclear magneton)

| **Baryons** **($J^P = 3/2^+$)** | **Effective mass scheme** |
|---|---|
| Decuplet $C = 0$ | |
| $\Delta^{++}$ | $3\mu_u^{eff}$ |
| $\Delta^{+}$ | $(2\mu_u^{eff} + \mu_d^{eff})$ |
| $\Delta^{0}$ | $(2\mu_d^{eff} + \mu_u^{eff})$ |
| $\Delta^{-}$ | $3\mu_d^{eff}$ |
| $\Sigma^{*+}$ | $(2\mu_u^{eff} + \mu_s^{eff})$ |
| $\Sigma^{*0}$ | $(\mu_u^{eff} + \mu_d^{eff} + \mu_s^{eff})$ |
| $\Sigma^{*-}$ | $(2\mu_d^{eff} + \mu_s^{eff})$ |
| $\Xi^{*-}$ | $(2\mu_s^{eff} + \mu_u^{eff})$ |
| $\Xi^{*0}$ | $(2\mu_s^{eff} + \mu_d^{eff})$ |
| $\Omega^{-}$ | $3\mu_s^{eff}$ |
| Sextet C = 1 | |
| $\Sigma_c^{*++}$ | $(2\mu_u^{eff} + \mu_c^{eff})$ |
| $\Sigma_c^{*+}$ | $(\mu_u^{eff} + \mu_d^{eff} + \mu_c^{eff})$ |
| $\Sigma_c^{*0}$ | $(2\mu_d^{eff} + \mu_c^{eff})$ |
| $\Xi_c^{*+}$ | $(\mu_u^{eff} + \mu_s^{eff} + \mu_c^{eff})$ |
| $\Xi_c^{*0}$ | $(\mu_d^{eff} + \mu_s^{eff} + \mu_c^{eff})$ |
| $\Omega_c^{*0}$ | $(2\mu_s^{eff} + \mu_c^{eff})$ |
| Triplet $C = 2$ | |
| $\Xi_{cc}^{*++}$ | $(2\mu_c^{eff} + \mu_u^{eff})$ |
| $\Xi_{cc}^{*+}$ | $(2\mu_c^{eff} + \mu_d^{eff})$ |
| $\Omega_{cc}^{*+}$ | $(2\mu_c^{eff} + \mu_s^{eff})$ |
| Singlet $C = 3$ | |
| $\Omega_{ccc}^{*++}$ | $3\mu_c^{eff}$ |

Table 2(a). $\left|\mu_{3/2^+\to 1/2^+}\right|$ Transition magnetic moments of charmless baryons using effective quark masses (in nuclear magneton)

| $3/2^+ \to 1/2^+$ **Transitions** | **Expressions** |
|---|---|
| $(C=0)$ *Decuplet* $\longrightarrow$ *octet* | |
| $\mu_{\Delta^+\to p}$ | $\frac{2\sqrt{2}}{3}(\mu_u^{eff} - \mu_d^{eff})$ |
| $\mu_{\Delta^0\to n}$ | $\frac{2\sqrt{2}}{3}(\mu_d^{eff} - \mu_u^{eff})$ |
| $\mu_{\Sigma^{*+}\to\Sigma^+}$ | $\frac{2\sqrt{2}}{3}(\mu_u^{eff} - \mu_s^{eff})$ |
| $\mu_{\Sigma^{*0}\to\Sigma^0}$ | $\frac{\sqrt{2}}{3}(2\mu_s^{eff} - \mu_u^{eff} - \mu_d^{eff})$ |
| $\mu_{\Sigma^{*0}\to\Lambda^0}$ | $\sqrt{\frac{2}{3}}(\mu_u^{eff} - \mu_d^{eff})$ |
| $\mu_{\Sigma^{*-}\to\Sigma^-}$ | $\frac{2\sqrt{2}}{3}(\mu_s^{eff} - \mu_d^{eff})$ |
| $\mu_{\Xi^{*0}\to\Xi^0}$ | $\frac{2\sqrt{2}}{3}(\mu_u^{eff} - \mu_s^{eff})$ |
| $\mu_{\Xi^{*-}\to\Xi^-}$ | $\frac{2\sqrt{2}}{3}(2\mu_s^{eff} - \mu_d^{eff})$ |

Table 2(b). Expressions for $\left|\mu_{3/2^+\to 1/2^+}\right|$ transition magnetic moments of charmed baryons using effective quark masses (in nuclear magneton)

| $3/2^+ \to 1/2^+$ **Transitions** | **Expressions** |
|---|---|
| (C = 1) *Sextet* $\to$ *anti-triplet* | |
| $\mu_{\Sigma_c^+\to\Lambda_c^+}$ | $\sqrt{\frac{2}{3}}(\mu_u^{eff}-\mu_d^{eff})$ |
| $\mu_{\Xi_c^{*+}\to\Xi_c^+}$ | $\sqrt{\frac{2}{3}}(\mu_u^{eff}-\mu_s^{eff})$ |
| $\mu_{\Xi_c^{*0}\to\Xi_c^0}$ | $\sqrt{\frac{2}{3}}(\mu_d^{eff}-\mu_s^{eff})$ |
| (C = 1) *Sextet* $\to$ *sextet* | |
| $\mu_{\Sigma_c^{*++}\to\Sigma_c^{++}}$ | $\frac{2\sqrt{2}}{3}(\mu_c^{eff}-\mu_u^{eff})$ |
| $\mu_{\Sigma_c^{*+}\to\Sigma_c^+}$ | $\frac{\sqrt{2}}{3}(2\mu_c^{eff}-\mu_u^{eff}-\mu_d^{eff})$ |
| $\mu_{\Sigma_c^{*0}\to\Sigma_c^0}$ | $\frac{2\sqrt{2}}{3}(\mu_c^{eff}-\mu_d^{eff})$ |
| $\mu_{\Xi_c^{*+}\to\Xi_c^{'+}}$ | $\frac{\sqrt{2}}{3}(2\mu_c^{eff}-\mu_u^{eff}-\mu_s^{eff})$ |
| $\mu_{\Xi_c^{*0}\to\Xi_c^{'0}}$ | $\frac{\sqrt{2}}{3}(2\mu_c^{eff}-\mu_d^{eff}-\mu_s^{eff})$ |
| $\mu_{\Omega_c^{*0}\to\Omega_c^0}$ | $\frac{2\sqrt{2}}{3}(\mu_c^{eff}-\mu_s^{eff})$ |
| (C = 2) *Triplet* $\to$ *triplet* | |
| $\mu_{\Omega_{cc}^{*0}\to\Omega_{cc}^0}$ | $\frac{2\sqrt{2}}{3}(\mu_s^{eff}-\mu_c^{eff})$ |
| $\mu_{\Xi_{cc}^{*++}\to\Xi_{cc}^{++}}$ | $\frac{2\sqrt{2}}{3}(\mu_u^{eff}-\mu_c^{eff})$ |
| $\mu_{\Xi_{cc}^{*+}\to\Xi_{cc}^+}$ | $\frac{2\sqrt{2}}{3}(\mu_d^{eff}-\mu_c^{eff})$ |

Table 3(a). Magnetic Moments of ($J^P = 3/2^+$) decuplet baryons using effective quark masses (in nuclear magneton)

| Baryons ($J^P = 3/2^+$) | Naïve Quark Model | Effective mass scheme | Screening Effect Scheme | Expt. | [28] | [36] | [18] NRQM | [18] ChQM |
|---|---|---|---|---|---|---|---|---|
| Decuplet $C = 0$ | | | | | | | | |
| $\Delta^{++}$ | 5.58 | 4.56 | 4.68 | 3.7~7.5[7]<br>4.52 ± 0.95 [8] | 4.52 | 5.39 | 5.43 | 5.30 |
| $\Delta^{+}$ | 2.79 | 2.28 | 2.36 | --- | 2.12 | 2.66 | 2.72 | 2.58 |
| $\Delta^{0}$ | 0 | 0 | -0.025 | --- | -0.29 | -0.08 | 0 | -0.13 |
| $\Delta^{-}$ | -2.79 | -2.28 | -2.34 | --- | -2.69 | -2.82 | -2.72 | -2.85 |
| $\Sigma^{*+}$ | 3.11 | 2.56 | 2.54 | --- | 2.63 | 2.82 | 3.02 | 2.88 |
| $\Sigma^{*0}$ | 0.32 | 0.23 | 0.14 | --- | 0.08 | 0.13 | 0.30 | 0.17 |
| $\Sigma^{*-}$ | -2.48 | -2.10 | -2.19 | --- | -2.48 | -2.56 | -2.41 | -2.55 |
| $\Xi^{*-}$ | -2.16 | -1.90 | -2.05 | --- | -2.27 | -2.30 | -2.11 | -2.25 |
| $\Xi^{*0}$ | 0.63 | 0.48 | 0.32 | -- | 0.44 | 0.34 | 0.60 | 0.47 |
| $\Omega^{-}$ | -1.84 | -1.67 | -1.90 | -1.94 ± 0.31 [9]<br>-2.02 ± 0.06 [10] | -2.06 | -2.05 | -1.81 | -1.95 |

Table 3(b). Magnetic Moments of charmed ($J^P = 3/2^+$) baryons using effective quark masses (in nuclear magneton)

| **Baryons** | **Naïve Quark Model** | **Effective mass scheme** | **Screening Effect Scheme** | **[33]** | **[29]** | **[35]** |
|---|---|---|---|---|---|---|
| Sextet C = 1 | | | | | | |
| $\Sigma_c^{*++}$ | 4.09 | 3.56 | 3.63 | 3.68 ~ 3.84 | 4.81±1.22 | -- |
| $\Sigma_c^{*+}$ | 1.30 | 1.17 | 1.18 | 1.20 ~ 1.26 | 2.00±0.46 | -- |
| $\Sigma_c^{*0}$ | -1.49 | -1.23 | -1.18 | -0.83 ~ -0.85 | -0.80±0.20 | -- |
| $\Xi_c^{*+}$ | 1.61 | 1.43 | 1.39 | 1.45 ~ 1.52 | -- | -- |
| $\Xi_c^{*0}$ | -1.18 | -1.00 | -1.02 | -0.67 ~ -0.69 | -- | -- |
| $\Omega_c^{*0}$ | -0.80 | -0.77 | -0.84 | -0.83 ~ -0.87 | -- | -- |
| Triplet $C = 2$ | | | | | | |
| $\Xi_{cc}^{*++}$ | 2.59 | 2.41 | 2.52 | 2.66 ~ 2.76 | -- | 2.67 |
| $\Xi_{cc}^{*+}$ | -0.20 | -0.11 | 0.035 | -0.16 ~ 0.17 | -- | -0.31 |
| $\Omega_{cc}^{*+}$ | 0.12 | 0.16 | 0.21 | 0.12 | -- | 0.14 |
| Singlet $C = 3$ | | | | | | |
| $\Omega_{ccc}^{*++}$ | 1.10 | 1.10 | 1.16 | 1.15~1.19 [34] | -- | -- |

Table 4(a). $\left|\mu_{3/2^+\to 1/2^+}\right|$ transition magnetic moments of charmless baryons using effective masses (in nuclear magneton)

| $3/2^+ \to 1/2^+$ **Transitions** | **Naïve Quark model** | **Effective mass scheme** | **Screening Effect Scheme** | **[28]** |
|---|---|---|---|---|
| (C=0) *Decuplet* $\to$ *octet* | | | | |
| $\mu_{\Delta^+\to p}$ | 2.63 | 2.48 | 2.33 | 2.76 |
| $\mu_{\Delta^0\to n}$ | 2.63 | 2.58 | 2.44 | 2.76 |
| $\mu_{\Sigma^{*+}\to\Sigma^+}$ | 2.33 | 2.13 | 2.10 | 2.24 |
| $\mu_{\Sigma^{*0}\to\Sigma^0}$ | 1.02 | 0.96 | 0.97 | 1.01 |
| $\mu_{\Sigma^{*0}\to\Lambda^0}$ | 2.28 | 2.25 | 2.12 | 2.46 |
| $\mu_{\Sigma^{*-}\to\Sigma^-}$ | 0.30 | 0.22 | 0.15 | 0.22 |
| $\mu_{\Xi^{*0}\to\Xi^0}$ | 2.33 | 2.27 | 2.21 | 2.46 |
| $\mu_{\Xi^{*-}\to\Xi^-}$ | 0.30 | 0.32 | 0.27 | 0.27 |

Table 4(b). $\left|\mu_{3/2^+\to 1/2^+}\right|$ transition magnetic moments of charmed baryons using effective masses (in nuclear magneton)

| $3/2^+ \to 1/2^+$ **Transitions** | **Naïve Quark Model** | **Effective mass scheme** | **Screening Effect Scheme** | **[30]*** |
|---|---|---|---|---|
| (C=1) *Sextet* $\to$ *anti-tiplet* | | | | |
| $\mu_{\Sigma_c^+\to\Lambda_c^+}$ | 2.28 | 2.28 | 2.15 | 1.48±0.55 |
| $\mu_{\Xi_c^{*+}\to\Xi_c^+}$ | 2.02 | 1.96 | 1.94 | 1.47±0.66 |
| $\mu_{\Xi_c^{*0}\to\Xi_c^0}$ | 0.26 | 0.25 | 0.18 | 0.16±0.07 |
| (C=1) *Sextet* $\to$ *sextet* | | | | |
| $\mu_{\Sigma_c^{*++}\to\Sigma_c^{++}}$ | 1.41 | 1.19 | 1.23 | 1.06±0.38 |
| $\mu_{\Sigma_c^{*+}\to\Sigma_c^+}$ | 0.09 | 0.04 | 0.08 | 0.45±0.11 |
| $\mu_{\Sigma_c^{*0}\to\Sigma_c^0}$ | 1.22 | 1.11 | 1.07 | 0.19±0.08 |
| $\mu_{\Xi_c^{*+}\to\Xi_c^{'+}}$ | 0.24 | 0.17 | 0.17 | |
| $\mu_{\Xi_c^{*0}\to\Xi_c^{'0}}$ | 1.07 | 1.00 | 0.99 | |
| $\mu_{\Omega_c^{*0}\to\Omega_c^0}$ | 0.92 | 0.88 | 0.90 | |
| (C=1) *Triplet* $\to$ *Triplet* | | | | |
| $\mu_{\Omega_{cc}^{*0}\to\Omega_{cc}^0}$ | 0.92 | 0.90 | 0.88 | |
| $\mu_{\Xi_{cc}^{*++}\to\Xi_{cc}^{++}}$ | 1.41 | 1.31 | 1.35 | |
| $\mu_{\Xi_{cc}^{*+}\to\Xi_{cc}^+}$ | 1.22 | 1.17 | 1.06 | |

*T.M. Aliev, K. Azizi and A. Ozpineci have given their results in natural magneton $(e\hbar/2cM_B)$, however to convert to nuclear magneton we multiply the entire magnetic moments with $2m_N/(M_{B_{3/2^+}}+M_{B_{1/2^+}})$.